\documentclass{article}
\usepackage{graphicx}  
\usepackage{color}
\usepackage{xcolor}
\usepackage{dcolumn}   
\usepackage{lineno}
\usepackage{bm}        
\usepackage{amssymb}   
\usepackage{amsmath,textcomp}
\usepackage[font=small,skip=0pt]{caption}
\usepackage{xfrac,ulem}
\usepackage[utf8]{inputenc}
\usepackage{float}
\usepackage{authblk}
\usepackage{hyperref}
\usepackage[margin = 1in]{geometry}
\usepackage{lineno}
\usepackage{afterpage}
\usepackage{pdfpages}

\hypersetup{%
    pdfborder = {0 0 0}
} 

\title{Surpassing the Standard Quantum Limit using an Optical Spring}
\author[1,+,*]{Torrey Cullen}
\author[1]{Ron Pagano}
\author[1]{Scott Aronson}
\author[1,**]{Jonathan Cripe}
\author[2]{Sarah Safura Sharif}
\author[1]{Michelle Lollie}
\author[1]{Henry Cain}
\author[3]{Paula Heu}
\author[3,5]{David Follman}
\author[3,4,5]{Garrett D. Cole}
\author[6]{Nancy Aggarwal}
\author[1,*]{Thomas Corbitt}
\affil[1]{Department of Physics \& Astronomy, Louisiana State University, Baton Rouge, LA, 70803}
\affil[2]{School of Electrical and Computer Engineering, University of Oklahoma, Norman, OK}\affil[3]{Crystalline Mirror Solutions LLC and GmbH, Santa Barbara, CA, and Vienna, Austria}
\affil[4]{Vienna Center for Quantum Science and Technology (VCQ), Faculty of Physics, University of Vienna, A-1090 Vienna, Austria}
\affil[*]{email: tcullen@caltech.edu; tcorbitt@phys.lsu.edu}
\affil[5]{Present address: Thorlabs Crystalline
Solutions, Santa Barbara, CA, USA}
\affil[6]{Northwestern University Department of Physics and Astronomy, Evanston, IL}
\affil[**]{Current Affiliation: Laboratory for Physical Sciences, College Park, MD 20740, USA }
\affil[+]{Current Affiliation: California Institute of Technology - The Division of Physics, Mathematics and Astronomy 1200 E California Blvd, Pasadena CA 91125}
\date{\today}

\begin{document}

\maketitle

\textbf{Quantum mechanics places noise limits and sensitivity restrictions on physical measurements. The balance between unwanted backaction and the precision of optical measurements imposes a standard quantum limit (SQL) on interferometric systems. In order to realize a sensitivity below the SQL, it is necessary to leverage a back-action evading measurement technique, reduce thermal noise to below the level of back-action, and exploit cancellations of any excess noise contributions at the detector. Many proof of principle experiments have been performed, but only recently has an experiment achieved sensitivity below the SQL. In this work, we extend that initial demonstration and realize sub-SQL sensitivity nearly two times better than previous measurements, and with an architecture applicable to interferometric gravitational wave detectors. In fact, this technique is directly applicable to Advanced LIGO, which could observe similar effects with a detuned signal recycling cavity. We measure a total sensitivity below the SQL by $\textbf{2.8}$ dB, corresponding to a reduction in the noise power by $\textbf{72}\pm\textbf{5.1}$ \% below the quantum limit. Through the use of a detuned cavity and the optical spring effect, this noise reduction is tunable, allowing us to choose the desired range of frequencies that fall below the SQL. This result demonstrates access to sensitivities well below the SQL at frequencies applicable to LIGO, with the potential to extend the reach of gravitational wave detectors further into the universe. }
\section{Introduction}
The standard quantum limit (SQL) is a theoretical limit imposed on precision measurements by the Heisenberg uncertainty principle\cite{Braginsky547,Chen_2013}. While its name implies an ultimate limit on the precision measurements, this limit can in fact be beaten through clever manipulations of the system under test. Ground based gravitational wave detectors such as LIGO have achieved sensitivities that approach the SQL at frequencies near 100 Hz \cite{2001PhRvD..65b2002K,2003PhRvD..67h2001B}. As of the conclusion of the third observing run, the aLIGO detectors are within a factor of 2-3 of the SQL at 70 Hz \cite{PhysRevD.102.062003,ligosql}, with this quantity expected to improve with the implementation of frequency dependent squeezing \cite{FDS_in_prep}. In this letter, we experimentally demonstrate a technique that allows interferometric gravitational wave detectors to reach sensitivities below the SQL. 

In an interferometric measurement such as LIGO, quantum noise exists in two parts: shot noise (imprecision noise) and quantum backaction (radiation pressure noise). Shot noise scales inversely with laser power whereas radiation pressure noise (RPN) scales proportionally to laser power. The uncertainty principle for these two quantities is given by
\begin{equation}
    S^{\mathrm{imp}}S^{\mathrm{rpn}}\geq \hbar^2/4 
\end{equation}
The SQL may be derived by analyzing this relationship. A spectral density for the imprecision and radiation pressure noise can be given by \cite{SQL_Schliesser,2014RvMP...86.1391A}
\begin{equation}
    S^{\mathrm{imp}} = \frac{x^2_{\mathrm{zpf}}}{4\Gamma_{\mathrm{meas}}}
    \label{eq:imp}
\end{equation}
\begin{equation}
    S^{\mathrm{rpn}} =\frac{\hbar^2 \Gamma_{\mathrm{meas}}}{x^2_{\mathrm{zpf}}}
    \label{eq:rpn}
\end{equation}
where $\Gamma_{\mathrm{meas}}$ is a measurement rate and $x_{\mathrm{zpf}}$ is the RMS of the oscillator's zero-point fluctuations \cite{SQL_Schliesser,2014RvMP...86.1391A}. 
By analyzing the equation of motion for a damped harmonic oscillator the mechanical susceptibility of the system is
\begin{equation}
    \frac{\Tilde{X}}{\Tilde{F}} = \chi(\Omega) = \frac{1}{m(\Omega^2_0-\Omega^2-i\Gamma_m\Omega)} ,
    \label{eq:suseptibility}
\end{equation}
where $\Gamma_m$ is a mechanical damping rate and $\Omega_0$ is the fundamental resonant frequency of the oscillator. Equations \ref{eq:imp}, \ref{eq:rpn}, and \ref{eq:suseptibility} lead to the total quantum noise (qn) in the system, given by
\begin{equation}
    S^{\mathrm{qn}}(\Omega) = S^{\mathrm{imp}} + |\chi(\Omega)|^2S^{\mathrm{rpn}}(\Omega) .
    \label{eq:S_quantum}
\end{equation}
This quantity can be minimized with respect to the measurement strength to determine the corresponding level of minimum noise,
\begin{equation}
    \Gamma^{\mathrm{opt}}(\Omega) = \frac{x^2_{zpf}}{2\hbar |\chi(\Omega)|} .
\end{equation}
Plugging this into Eq \ref{eq:S_quantum} yields what is known as the SQL,
\begin{equation}
    S^{\mathrm{qn}} = S_{\mathrm{SQL}} = \hbar |\chi(\Omega)|.
    \label{eq:SQL}
\end{equation}
The calculations of the SQL used in this letter makes use of Equation \ref{eq:SQL} through code developed in Ref. \cite{Corbitt_mathematical}. This calculation shows that a lower limit exists on the total quantum noise in a system whose imprecision noise and backaction remain uncorrelated. However, sub-SQL sensitivities can be achieved by correlating and manipulating these two noise sources in a process referred to as quantum nondemolition (QND) \cite{2001PhRvD..65b2002K,Braginskiĭ_1975}. 

A variety of proof-of-principle sub-SQL QND techniques have been explored \cite{Regal_variational,Suhdetecting,doi:10.1126/science.aac5138,PhysRevX.5.041037,MJYAP_squeezing}. One method involves injected squeezed light, something that was accomplished by the detector group at aLIGO. In that work, a measurement of up to $3$ dB below the SQL was realized by making use of injected squeezed vacuum states and by subtracting unwanted classical noise from the measurement \cite{ligosql}. A second technique also takes advantage of a variational readout which makes use of a second field, often originating from the same laser source, to modify the measurement quadrature \cite{2003PhRvD..67h2001B,ligosql}. This letter makes use of a third method, by amplifying the mirror's motion via the optical spring effect. This amplifies the back action and any signal resulting from the mirror's motion at the optical spring frequency while simultaneously keeping shot noise the same. A detailed calculation of why the mirror's motion is amplified can be found in the following section. Detailed calculations of the optical spring effect can be found in Refs \cite{2001PhRvD..64d2006B,2002PhRvD..65d2001B,2006PhRvD..73f2002K}. 

Not until recently has has it been possible to experimentally demonstrate any interferometric measurement with sensitivity below the SQL \cite{SQL_Schliesser,scienceSQL}. The Mason et al. experiment utilized a cryogenically cooled high quality factor Si$_3$N$_4$ membrane resonator dispersively coupled to a Fabry Perot cavity, which allowed a sensitivity measurement up to 1.5 dB below the SQL, at megahertz frequencies. We go beyond this previously ground breaking measurement by performing the experiment in the audio band, and forgoing dispersive coupling by instead employing a linear two-mirror cavity with a movable mirror. This operating range and optical setup is more akin to the current design of ground based gravitational wave detectors.


Previous experiments performed by our group \cite{nancy_squeeze,PhysRevX.10.031065} were carried out at room temperature. In order to surpass the SQL, two main upgrades to our system must be made. First, thermal noise must be combatted. This is realized with the addition of a cryostat cooler, bringing the cavity down to approximately $30$ K. Secondly, the frequency noise of the laser begins to limit noise levels at cryogenic temperatures. Frequency noise is addressed with the addition of a delay line interferometer. 

\section{Explanation of Optical Spring Suppression}
\label{section:Methods}

\label{app:OS}
Here we dive deeper into the physics behind why the optical spring allows for a sub-SQL operation. To understand the optical spring, consider the circulating power in a cavity given by, 
\begin{equation}
P_{C}=\frac{P_{0}}{1+\delta^{2}},
\label{eq:P_circ}
\end{equation}
where $P_0$ is the maximum circulating power with the cavity on resonance and $\delta$ is the detuning in terms of linewidths of the cavity. In the case of a cavity in which the only loss is the transmission through one mirror, in our case the input mirror in Figure \ref{fig:Setup_SQL}, the max circulating power is given by $P_0 = \frac{4}{T} P_{in}$, where $T$ mirror transmission. The force due to radiation pressure on one mirror is then $\frac{2P_c}{c}$. The optical spring constant can be found by taking the derivative of this with respect to $x$,
\begin{equation}
\begin{aligned}
K_{O S} =\frac{2}{c} \frac{d P_{c}}{d x} 
 =\frac{2}{c} \frac{d P_{c}}{d \delta} \frac{d \delta}{d x} 
& =\frac{-32 \pi \delta P_{c}}{\lambda c T\left(1+\delta^{2}\right)} .
\end{aligned}
\label{eq:optical_spring_constant}
\end{equation}
Note that from Eq \ref{eq:optical_spring_constant}, a cavity detuning of $\delta < 0$ corresponds to a positive optical spring constant and therefore a restoring force. 

All motion of the movable mirror in the cavity will be amplified at the optical spring resonance frequency, given by $\Omega_{OS} = \sqrt{\frac{K_{OS}+K_m}{m}}$, where $K_m$ is the mechanical spring constant and $m$ is the mass of the movable mirror. For this experiment, $K_{OS} \gg K_m$, and so we will approximate $\Omega_{OS} \approx \sqrt{\frac{K_{OS}}{m}}$. This calculation does not take into account the time delay in the cavity’s response that gives rise to an optical damping force $\Gamma_{OS}$. Additionally, this calculation assumes the mechanical damping rate is very small compared to the optical damping force, $\Gamma_m \ll \Gamma_{OS}$. For our parameters, $\Gamma_{OS} \ll \Omega_{OS}$, and therefore the magnitude of the quality factor of the optical spring is large, $Q_{OS} \gg 1$. This implies that the mirror motion, including back action, is amplified at the optical spring resonance compared to the absence of the optical spring in a similar system.  Effectively, at the resonance, the back action and any signal resulting from mirror motion are amplified by a factor of $Q_{OS}$, while the shot noise remains the same. In the context of gravitational wave detection, sensitivity is ordinarily referred to free mass displacement, or the equivalent displacement sensitivity if the system were a free mass. This is done because the strain sensitivity may then be obtained simply by dividing by the length of the cavity. At the resonance of the optical spring, this is accomplished by dividing by $Q_{OS}$ to obtain the original, unamplified motion. In doing so, the shot noise is reduced by the same factor $Q_{OS}$. Essentially, this results in a linear rescaling, such that in the optical spring picture, all motion, including back action, is amplified by $Q_{OS}$ and shot noise remains unchanged. In the free mass picture, all motion, including back action, remains the same, but shot noise is reduced by $Q_{OS}$. Thus, in evaluating the performance at $\Omega_{OS}$ in the free mass picture, it is sufficient to consider only back action noise and neglect shot noise. 

To calculate the back action, we consider the radiation pressure imposed by the fluctuations in the circulating power of the cavity imposed by the shot noise of the incident light to the cavity,
\begin{equation}
\begin{aligned}
x_{r p} & =\frac{1}{m \Omega^{2}} \frac{2 P_{c}}{c} \sqrt{\frac{2 h f}{P_{i n}}} \\
& =\frac{1}{m \Omega^{2}} \frac{2 P_{c}}{c} \sqrt{\frac{8 h f}{P_{c} T\left(1+\delta^{2}\right)}}.
\end{aligned}
\end{equation}
To evaluate the performance, we can divide this by the free mass SQL,
\begin{equation}
\begin{aligned}
x_{S Q L} & =\sqrt{\frac{2 \hbar}{m \Omega^{2}}} \\
\end{aligned}
\end{equation}
and evaluate at the optical spring frequency $\Omega_{OS}$,
\begin{equation}
\begin{aligned}
\left. \frac{x_{r p}}{x_{S Q L}}\right|_{\Omega_{OS}} & =\frac{1}{m \Omega_{O S}^{2}} \frac{2 P_{c}}{c} \sqrt{\frac{8 h f}{P_{c} T\left(1+\delta^{2}\right)}} \times \sqrt{\frac{m \Omega_{O S}^{2}}{2 \hbar}} \\
& =\frac{1}{c} \sqrt{\frac{32 \pi h f P_{c}}{m \Omega_{O S}^{2} T\left(1+\delta^{2}\right)}} \\
\end{aligned}
\end{equation}
By applying the approximation for the optical spring frequency, as given above, we then end up with:
\begin{equation}
\begin{aligned}
\left. \frac{x_{r p}}{x_{S Q L}}\right|_{\Omega_{OS}} &=\frac{1}{c} \sqrt{\frac{32 \pi h f P_{c}}{m \left(\sqrt{\frac{K_{OS}}{m}} \right)^2 T\left(1+\delta^{2}\right)}}\\
&=\frac{1}{c} \sqrt{\frac{\lambda f c}{-\delta}} \\
&=\sqrt{\frac{1}{-\delta}} .
\end{aligned}
\end{equation}

Here we can see that this method, measuring at the optical spring resonance in transmission, beats the SQL by a factor of $\sqrt{-\delta}$. Note that a negative detuning corresponds to a positive optical spring and therefore a restoring force. This method is particularly robust against optical losses, and it relies on amplifying the signal rather than canceling out noise terms. This result is confirmed by numerical calculation. For a cavity with positive detunings, and therefore negative spring constant, the motion is not amplified and this calculation does not apply. 

\section{Experimental Setup and Procedures}
Our experimental setup consists of three main subsystems: the optomechanical (OM) cavity and associated feedback, a $100$ meter delay line interferometer, and an intensity stabilization servo (ISS). The OM cavity utilizes a 1-cm long Fabry-Perot configuration incorporating a half inch input mirror with a radius of curvature of $2.5$ cm and a second $70$ $\mathrm{\mu}$m diameter GaAs/AlGaAs (23 pairs) Bragg reflector integrated on a low-noise single-crystal GaAs cantilever \cite{nancy_squeeze,Cripe_QRPN}. The suspended micromirror has a mass of $50$ ng and a fundamental resonance frequency of $876$ Hz. The OM cavity is housed in a vacuum chamber ($\approx 10^{-8}$ torr) and is cooled cryogenically to about $30$ K. Cavity readout is realized via a Nd:YAG non-planar ring oscillator (NPRO) laser with wavelength 1064 nm.

\begin{figure}
\center
\includegraphics[width= \linewidth]{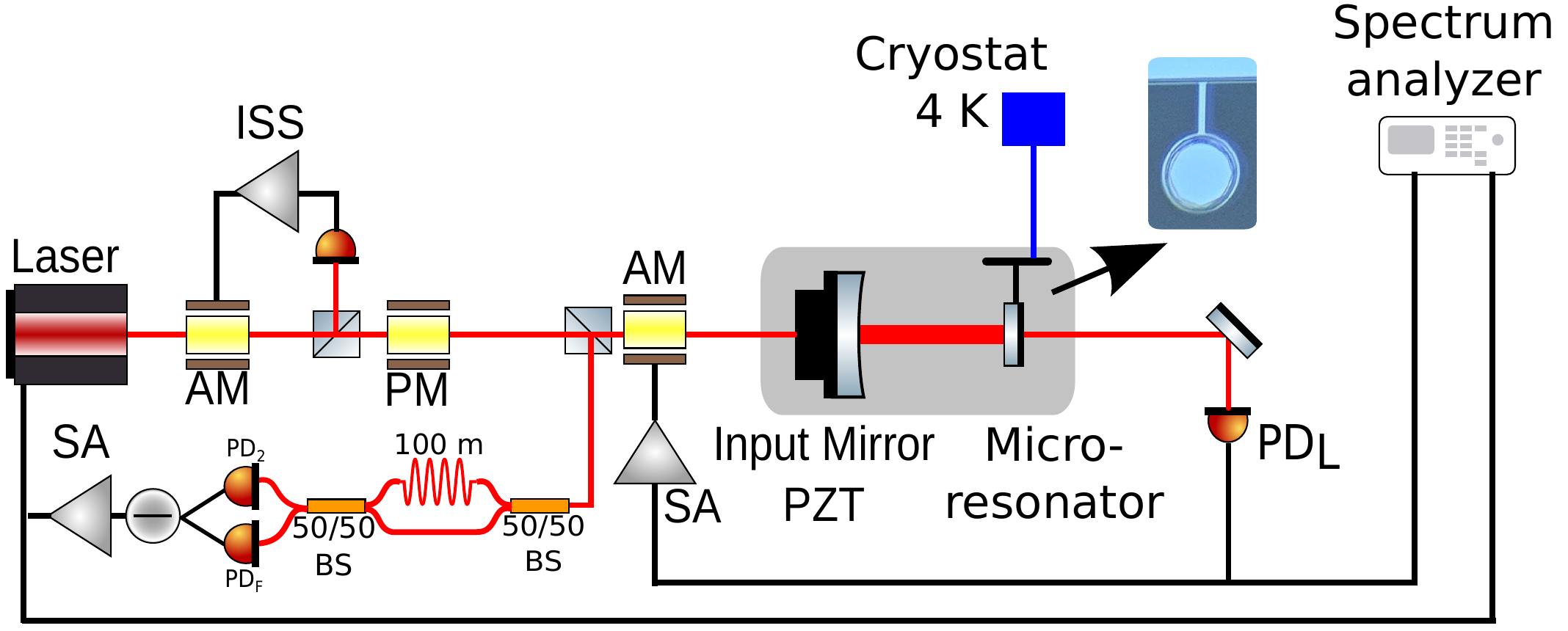}
\caption{\textbf{Overview of the experimental setup.}  
The microresonator is connected to a 4 K cryostat via a flexible heat link. Light exiting the cavity is measured on a photodetector ($PD_L$) where the signal is subsequently fed to a servo amplifier (SA) and to an Amplitude Modulator (AM), in order to stabilize the cavity at a constant detuning. An intensity stabilization servo (ISS) is used to stabilize the laser power to shot noise by feeding back to the first AM . Additionally, light is picked off before the cavity and fed to a 100 m delay line interferometer. This allows for a precise measurement of the frequency noise of the laser in order to subtract it from our final measurement.}
\label{fig:Setup_SQL}
\end{figure}

The delay line interferometer is created by a free space path and the addition of a 100 m long optical fiber that creates the optical path length difference between the two arms. The entirety of the delay line is kept at room temperature and atmospheric pressure.  This subsystem has two functions. First, it is used to suppress the laser frequency noise (LFN). The interference of the two paths in the delay line is measured on two photodetectors, $PD_F$ and $PD_2$ in Fig. \ref{fig:Setup_SQL}, and the two signals are subtracted from one another and sent to a PID controller to stabilize these fluctuations. This allows the interferometer to be locked to a stable value in order to facilitate its second function: precisely calibrating the LFN. With a precise calibration of the LFN and the measurement scheme described at length in the supplementary materials, we can study the noise spectrum free of LFN. Additionally, the final subsystem, an intensity stabilization servo (ISS), is used to suppresses classical relative intensity noise fluctuations. 

At room temperature, thermal noise is much larger than the SQL, so we introduce a two stage cryocooler. The minimum temperature reached on the sample holder is $16$ K. However, the large mechanical vibrations introduced by the cryostation compressor frequently result in a loss of the cavity lock. For this reason, the cryostation compressor is turned off and the cavity slowly warms as a measurement is performed. The recorded temperature refers to the final temperature reached at the end of the calibration and measurement process, thus yielding an upper bound on the theoretical estimate of the thermal noise.

To compare the measured noise to the SQL we make use of modified measurement scheme from previous experiments. The ISS and cryostation reduce classical intensity noise (CIN) and thermal noise below the SQL, but laser frequency noise still remains. A precise measurement of the LFN is taken concurrently with the displacement measurement, as well as the coherence between the two, in order to allow for a coherent subtraction of the LFN from the displacement measurement. 

The light transmitted through the optomechanical cavity is maximized on $\mathrm{PD_L}$. The signal from $\mathrm{PD_L}$ is sent through a servo amplifier (SA) before being sent to the second AM to lock the cavity. 
We lock the cavity with a detuning of $\delta = -3.1$. The linewidth of the cavity (half width half maximum) is $520$ kHz. Given an absolute detuning of 3.1 linewidths, the corresponding circulating power is $71$ mW and the optical spring frequency is $67$ kHz. The mechanical quality factor of the cantilever at cryogenic temperatures is $25000 \pm 2200$. 
\begin{figure}[!ht]
\center
\includegraphics[width=\linewidth]{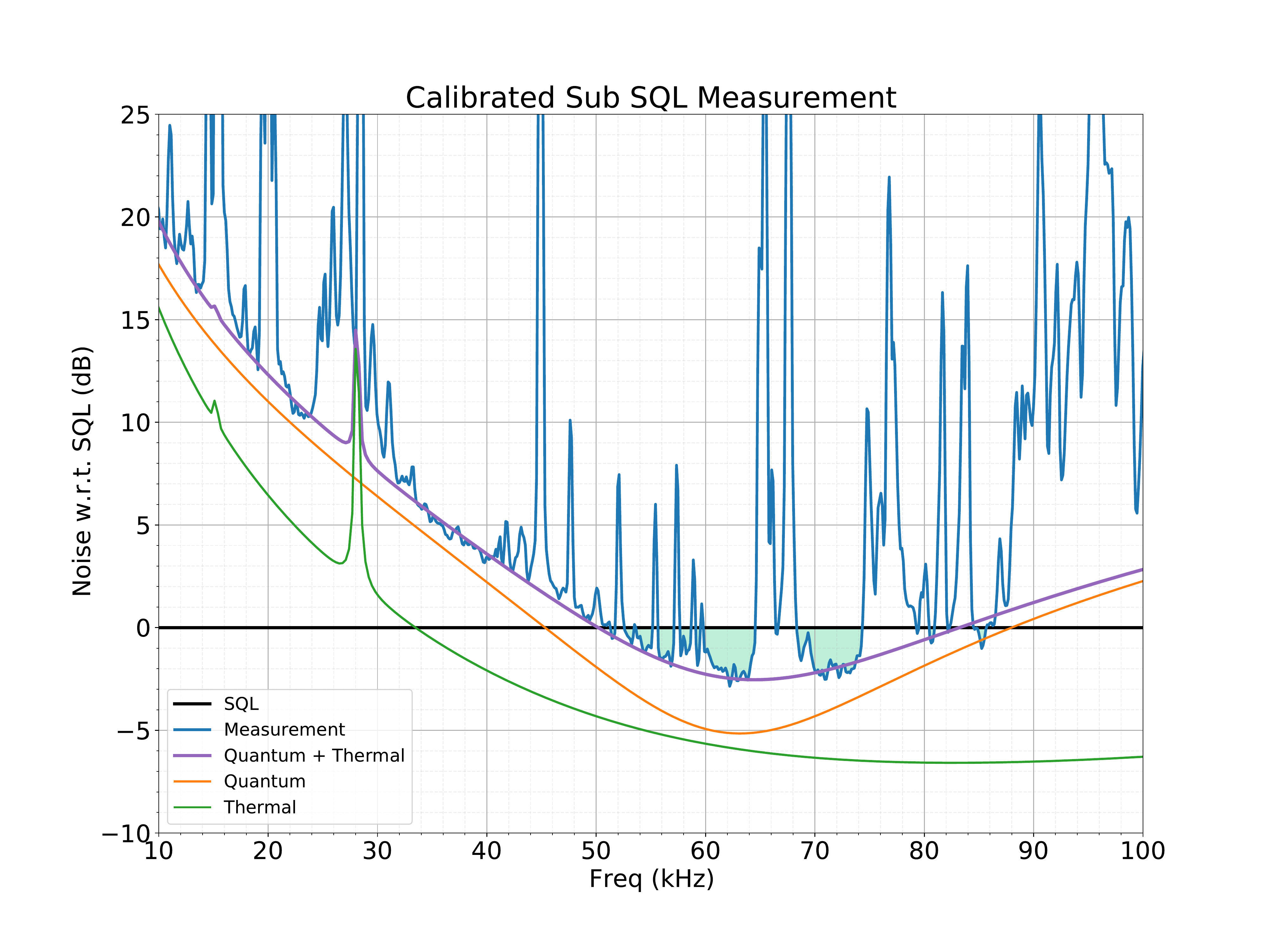}
\caption{\textbf{Calibrated noise spectrum with a $\textbf{67.8}$ kHz optical spring frequency.} Calibrated noise measurement as a ratio with respect to the standard quantum limit (blue). The orange and green curves represent the total quantum and thermal noise of the system respectively. The purple curve is the quadrature sum of the two, showing these two noise sources are the main contributions to the limiting noise of the experiment. This measurement realized a maximum $2.8$ dB or $72\%$ reduction below the SQL.} 
\label{fig:SQL1}
\end{figure}
\section{Results and discussion}

A displacement measurement as a ratio to the SQL of the system can found in Fig. \ref{fig:SQL1}. At an upper limit temperature of $29$ K, we observe up to a $2.8$ dB reduction below the SQL. Additionally, the frequency band at which the measurement lies below the SQL is between $50$ kHz - $74$ kHz, with the maximum reduction occurring at $62.2$ kHz. This maximum reduction corresponds to a $72\%$ reduction in the power spectral density. The orange and green curves represents the expected quantum and thermal noise respectively, as calculated by Ref \cite{Corbitt_mathematical}. Not pictured in Figure 2 is a measured noise level due to the electronics noise of the system, which we have found to be insignificant.

\begin{figure}[!ht]
\center
\includegraphics[width=\linewidth]{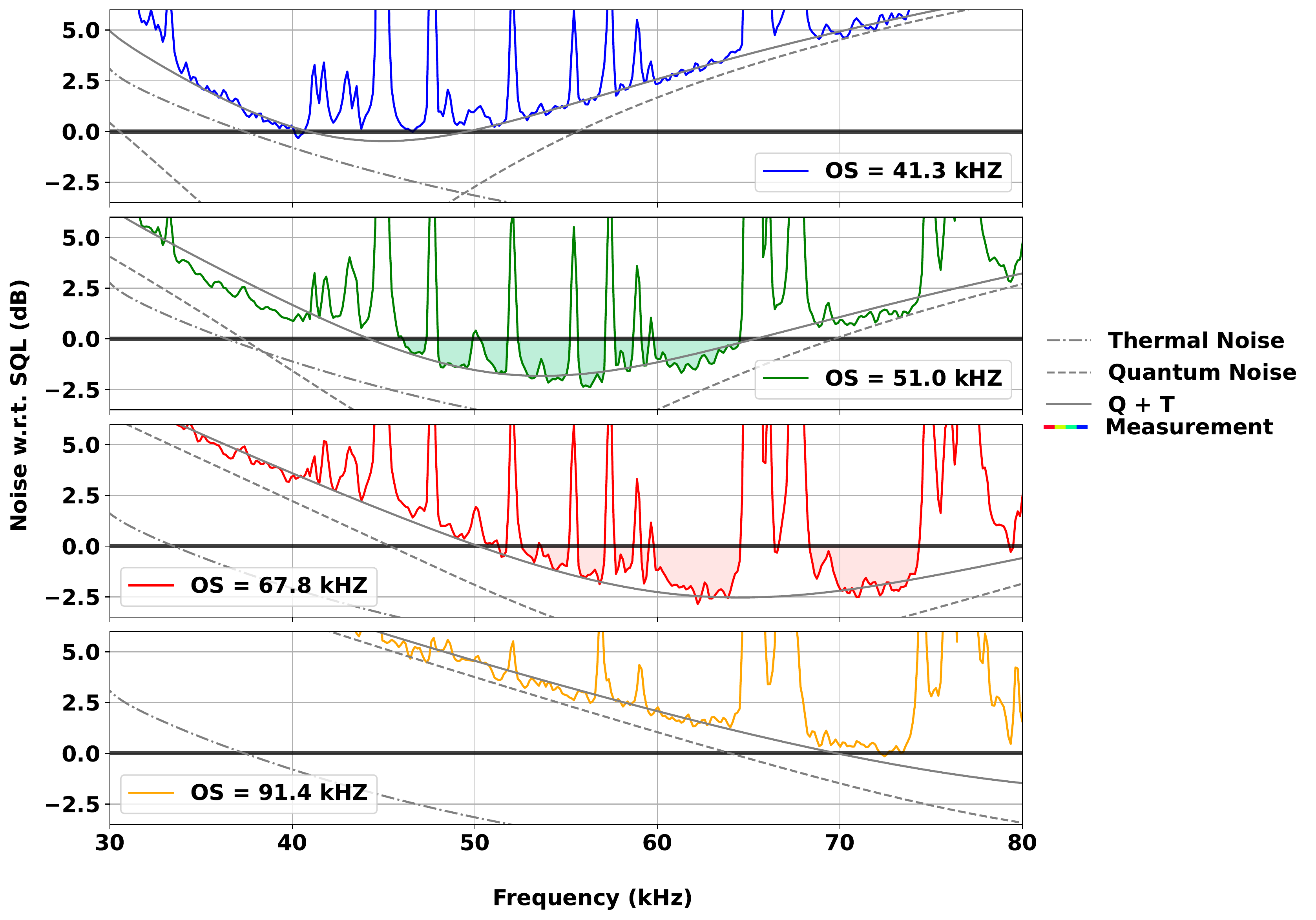}
\caption{\textbf{Calibrated measurements at four different optical spring frequencies.} Changing the detuning of the cavity via the optical spring effect tunes the frequency region at which we see a reduction below the SQL. $41.3$ kHz and $91.4$ kHz are the lower and upper limit, respectively, of the optical frequencies which yield an interferometric displacement measurement at or below the SQL. For optical spring frequencies between these values we measure some reduction below the SQL. In grey are calculated quantum and thermal noise corresponding to the parameters of each individual measurement.}
\label{fig:tuneable}
\end{figure}

As the optical spring generates amplification of the signal due to the mirror's motion, we can adjust the frequency at which the measurement goes below the SQL by changing the optical spring frequency. In this sense, we can create a tunable sub-SQL device as seen in Fig. \ref{fig:tuneable}. The optical spring frequency range that we observe sensitivities at or below the SQL is $41.3$ kHz - $91.4$ kHz. As seen in the top and bottom panels, at an optical spring frequency of $41.3$ and $91.4$ kHz, the calibrated noise measurement shown has the most sensitive portion equal to the SQL. For all optical spring frequencies in between, we see a reduction below the SQL. For the full range of optical spring strength, the measurement frequency range where sub-SQL sensitivity is observed is between $40$ kHz - $74$ kHz. Such tunability could be exploited for LIGO and other interferometric experiments if there is a specific frequency range of interest for realizing sub-SQL sensitivity.

An at length discussion of the calibration technique can be found in the Methods section. Here we provide details on the uncertainties associated with the experiment. There are two main contributions to the total experimental uncertainty: the uncertainty associated with our measure of the total noise, or our calibration technique, and the uncertainty in the magnitude of the SQL of our system, stemming from an uncertainty in the mass of the cantilever mirror.

For a robust sub-SQL claim, we require that the uncertainty in the calibration technique be much less than the amount we beat the SQL by. For this reason it is important to constrain an uncertainty in the calibration process. This can be done through testing the repeatability of the calibration process. We introduce a sine wave of a known voltage on the Electro-Optic Modulator (EOM), and measure the frequency response of this signal. This process was repeated to show that the measurement variation remained less than $1\%$ setting the maximum error in calibration at this level.

Additionally, we must take into consideration how well we know the SQL for our system. In the free mass regime, this solely depends on the mass of the micro-resonator mirror. In ref \cite{Cripe_QRPN}, the mass of the movable mirror was estimated to be $50$ ng. This was estimated through a calibrated measurement of the thermal noise of the system, in which Cripe et al. estimates a $\pm 10\%$ uncertainty in the mass. 
Propagating this through the SQL free mass formula corresponds to a $5\%$ uncertainty in the SQL of our system. If we consider the ratio of our calibrated measurement to the SQL as our final quantity, such as the blue curve in Fig. \ref{fig:SQL1}, the total relative error of the experiment is $5.1\%$. This error is significantly less than which we beat the SQL by.

The displacement sensitivity of this system has been shown to realize a long standing goal in the quantum optics community. The results presented here show that the standard quantum limit can be beaten at frequencies and using mechanical systems relevant to gravitational wave detectors. Additionally, the creation of a tunable sub-SQL device has the potential to enable improved gravitational wave detector sensitivity through the use of a detuned signal recycling cavity. 

\section{Laser frequency noise}
\indent This section describes the process we use to convert the measured signal on the photodetector, PD$_L$, initially in $\frac{V}{\sqrt{Hz}}$, to a displacement measurement in $\frac{m}{\sqrt{Hz}}$. This requires an intermediary step of first converting from $\frac{V}{\sqrt{Hz}}$ to $\frac{Hz}{\sqrt{Hz}}$ by using a Mach-Zender interferometer with a $100$ m delay line. This can be done by examining the electric field incident on the photodetector at the end of the delay line. The light arriving at either $PD_F$ and $PD_2$ is comprised of two contributions, one from each arm of the delay line. This can be written as,
\begin{equation}
    E = c_1 + c_2e^{i\omega L/c},
    \label{eq:efield_initial}
\end{equation}
where the first term is from the shorter arm, and the second term is due to the phase accumulated by the light in the longer arm. Because the difference between the length of the two arms (L) can be related to the time ($\tau$) the light takes to traverse the path, Equation \ref{eq:efield_initial} can be rewritten as,
\begin{equation}
    E = c_1 + c_2e^{i \omega \tau}.
\end{equation}
The photodetector measures the power incident on the photo detector, and then converts it to a proportional voltage with a negligible constant. This can be written as,
\begin{equation}
\begin{split}
    V =& |c_1 + c_2e^{i \omega \tau}|^2 \\
      =&A+B\mathrm{cos}(\omega \tau), 
\label{eq:split}
\end{split}
\end{equation}
where $c_1$ and $c_2$ have been rewritten in terms of new constants A and B. A and B represent the offset and amplitude of the interference pattern respectively. These quantities can be determined experimentally by driving the temperature of the monolithic laser crystal, which changes the frequency of the light produced by the laser. This process is also beneficial as it allows for an imperfect visibility when the two paths of the delay line interfere.

The derivative with respect to $\omega$ of Eq. \ref{eq:split} is then,
\begin{equation}
    \frac{dV}{d\omega} = -B\tau \mathrm{sin}(\omega \tau)  .
    \label{eq:dvdw}
\end{equation}
$\tau$ is obtained by sending a signal into a phase modulator (PM) and finding the frequency $f$ at which the signal on $PD_F$ is minimized. The second harmonic is used because the first happens to lie on the PM's, resonance making the signal noisy. Because of this $\tau$ is then,
\begin{equation}
    \tau = 2/f.
\end{equation}
We can then solve for the phase delay of the light, $\omega\tau$, using Equation \ref{eq:split},
\begin{equation}
    \omega\tau = \mathrm{cos}^{-1}\frac{V_L - A}{B},
\end{equation}
where $V_L$ is the average value of the offset on $PD_F$. This value is kept constant via a PID controller and is fed back to the frequency tuning input of the laser. Plugging this into Eq \ref{eq:dvdw} yields,
\begin{equation}
    \frac{dV}{d\omega} = -B\tau \mathrm{sin}(\mathrm{cos}^{-1}\frac{V_L - A}{B})  .
    \label{eq:dvdw_final}
\end{equation}
The quantity in Eq \ref{eq:dvdw_final} has units of $\frac{V}{Hz}$ and is used to convert the light exiting the cavity, measured at $PD_L$, from $\frac{V}{\sqrt{Hz}}$ to $\frac{Hz}{\sqrt{Hz}}$. 

Now that we have a quantity in $\frac{Hz}{\sqrt{Hz}}$, we can use the length of the cavity, L, and the wavelength of light in the cavity, $\lambda$, to convert to a displacement measurement. By multiplying the spectrum in $\frac{Hz}{\sqrt{Hz}}$ by $\frac{L}{c/\lambda}$, we obtain a quantity in $\frac{m}{\sqrt{Hz}}$. 

\section{Calibration and uncertainty}

In order study the light exiting the cavity without laser frequency noise, we measure the frequency noise directly and subtract it from the measurement. In this experiment, this is done by first considering the power spectrum of the two signals, one from the feedback locking photodetector, $PD_L$, and the other at the end of the delay line, $PD_F$,

\begin{equation}
    PD_L:\hspace{1mm} S_{11} = (F^2 + C^2_1)\lambda_1^2 
    \label{eq:pdf}
\end{equation}
\begin{equation}
    PD_F:\hspace{1mm}  S_{22} = F^2 + S_n^2
    \label{eq:pdl}
\end{equation}
where F is the frequency noise of the laser, $C_1$ is the signal from the transmitted light of the cavity, $\lambda_1$ is the factor used to represent the ratio of gains between the two photodetectors, and $S_n$ is the shot noise contribution of $PD_F$. The cross correlation of the two signals can be written as,
\begin{equation}
    S_{12} = \lambda_1F^2.
    \label{eq:cc}
\end{equation}
The coherence between the two signals is defined as,
\begin{equation}
    C = \frac{\langle S_{12} \rangle^2}{S_{11}S_{22}}.
    \label{eq:coherence}
\end{equation}

\begin{figure}[!ht]
    \centering
    \includegraphics[width=.75\linewidth]{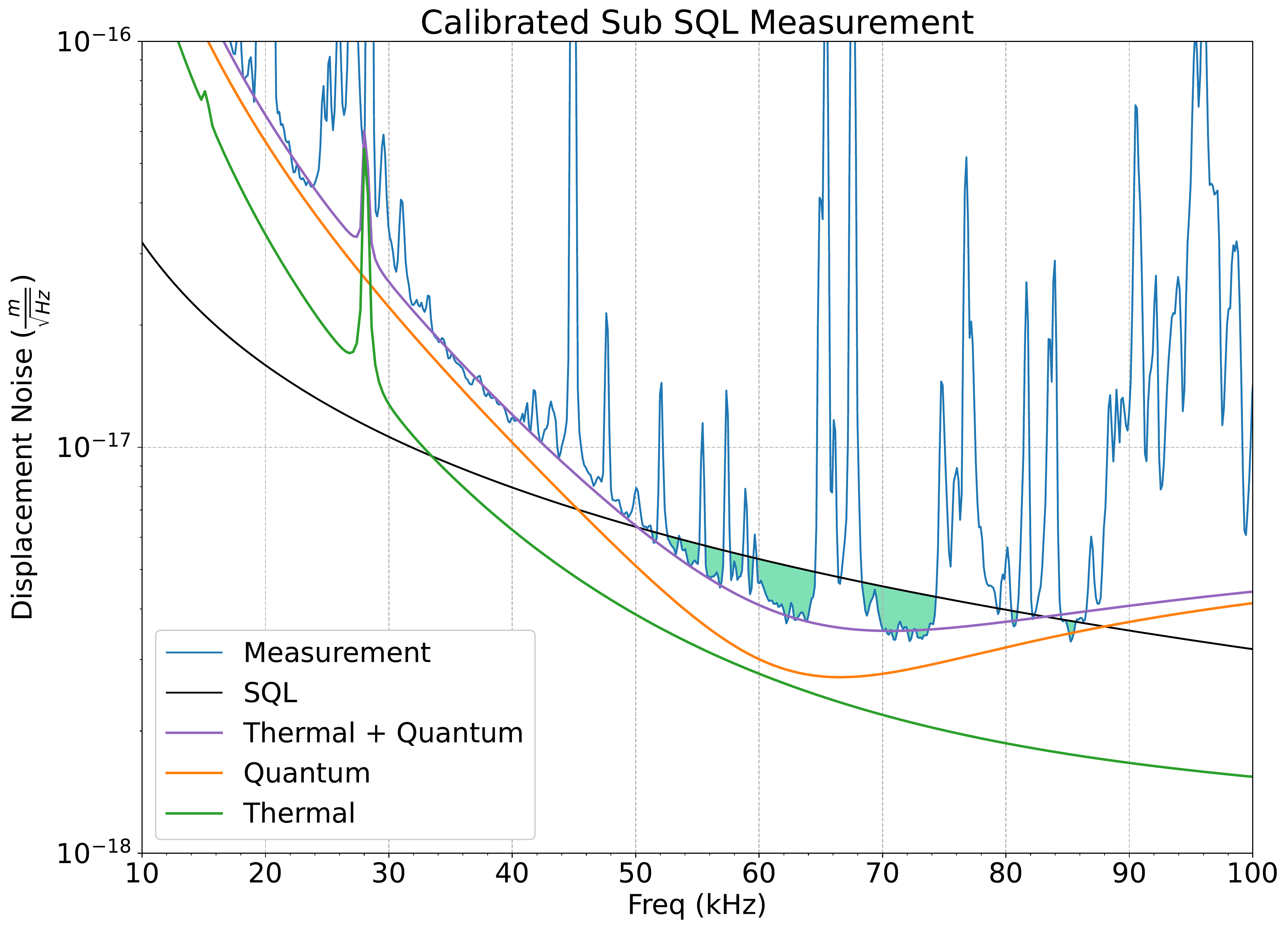}
    \caption{\textbf{Results of best sub-SQL measurement shown in the calibrated $\frac{m}{\sqrt{Hz}}$ units}. 67.8 kHz optical spring frequency measurement with 2.8 dB below the SQL shown in calibrated $\frac{m}{\sqrt{Hz}}$. This is compared to the SQL in the same units.}
    \label{fig:mainresult_meters}
\end{figure}

Plugging Equations \ref{eq:pdf}, \ref{eq:pdl}, \ref{eq:cc} into Equation \ref{eq:coherence}
\begin{align}
    C&=\frac{F^4}{(C_1^2+F^2)(F^2+S_n^2)}.
    \label{eq:updated_C}
\end{align}
Rearranging Equation \ref{eq:updated_C} for $C_1$, we find a frequency noise free measurement of the cavity can obtained by measuring the quantity,
\begin{equation}
    C_1 = F(\frac{F^2}{CS_{22}}-1)^\frac{1}{2}.
    \label{eq:C_1}
\end{equation}
Figure \ref{fig:mainresult_meters} shows the result of this calibration process compared to the SQL.

\bibliographystyle{ieeetr}
\bibliography{biblioRPL}
\clearpage

\section*{Acknowledgements}
T. Cullen, R.P, S.A, H.C., and T. Corbitt are supported by the National Science Foundation grants PHY-1150531 and PHY-1806634. NA is supported by NSF grant PHY--1806671 and a CIERA Postdoctoral Fellowship from the Center for Interdisciplinary Exploration and Research in Astrophysics at Northwestern University. A portion of this work was performed in the UCSB Nanofabrication Facility, an open access laboratory. This document has been assigned the LIGO document number LIGO-P2200263. The authors would also like to acknowledge Ian Macmillan for useful discussions.

\section*{Author contributions}
Torrey J Cullen (T.J.C.) led the work, with this project being a major focus on his doctoral dissertation. This work was supervised by Thomas R Corbitt (T.R.C.). J.C. and T.R.C. designed and built the optomechanical cavity and a majority of its subsystems. T.J.C., H.C., and T.R.C. designed and built the laser frequency noise detection system and calibration process. T.J.C., R.P., S.S.S, and T.R.C. performed the measurements and performed the data analysis. T.J.C. wrote the manuscript with help from R.P. and T.R.C., and very helpful discussions with J.C.,G.D.C., and N.A.. P.H., D.F., and G.D.C. fabricated the cantilever mirror used in this experiment.

\section*{Competing interests}
The authors declare no competing interests.
\clearpage

\end{document}